# Background for Terrestrial Antineutrino Investigations: Scientific Basis of Knowledge on the Composition of the Deep Interior of the Earth

J. Marvin Herndon

Transdyne Corporation
San Diego, California 92131 USA

July 12, 2004

## Abstract

I present from a historical perspective a logical progression of understanding, related to the composition of the deep interior of the Earth, that comes from fundamental discoveries and from discoveries of fundamental quantitative relationships in nature. By following step by step the reasoning from that understanding, one might begin to appreciate what is not yet known that pertains to recent interest in geo-antineutrinos and also what should be investigated to further advance that understanding.

## Introduction

Recent interest in the detection of geo-antineutrinos with spectral and directional resolution[1-3] is enthusiastically welcomed as a potential means of verifying the existence of a nuclear georeactor at the center of the Earth[4-6]. Generally, as physicists venture into a new area, there is a learning curve and there may be some confusion. But as physicists venture into the area of solid-Earth science, confusion may be considerably magnified by an Earth science literature turgid with decades of reports of investigations that fail to follow long-established, ethical standards of science. One major problem is that many investigators make models, based upon arbitrary assumptions, or based upon other models, which are themselves based upon arbitrary assumptions. Such models, sometimes designated by the makers as "reference model" or "preferred model", often ignore contradictory scientific evidence and can generally be replaced by different models based upon other assumptions, all of which may have questionable relevance, utility, and correctness.

The purpose of science is not to make arbitrary models based upon assumptions, but rather, to determine the true nature of the Earth and the Cosmos, which can be done by making fundamental discoveries and by discovering fundamental quantitative relationships in nature. In the following, I present from a historical perspective a logical progression of understanding about the Earth that comes from such fundamental discoveries and fundamental relationships. By following step by step the reasoning from that understanding, one might begin to realize the extent of what is not yet known pertaining to recent interest in geo-antineutrinos and what investigations should be undertaken to further advance that understanding.

**Historical Basis**

In 1897 Emil Wiechert[7] realized from the density determination of Henry Cavendish[8] that the Earth can not consist wholly of rock. Having seen in museums meteorites that consist entirely of nickeliferous iron metal, as well as meteorites made of both iron metal and stone, Wiechert suggested that the Earth has at its center a core, like the metallic iron of meteorites. The existence of such a core, Wiechert estimated, could account for the high bulk density of the Earth.

In 1906 Richard Oldham[9] determined the speed of earthquake-waves as a function of depth of travel within the Earth. He found generally that beneath the crust the velocities of earthquake-waves increase with increasing depth, but only to a particular depth, below which their velocities become abruptly and significantly slower. Oldham had discovered the Earth's core.

During the next twenty-five years, the dimension of the Earth's core was determined precisely and its state was shown to be liquid due to its failure to support transverse earthquake-waves[10]. Seismological data, augmented with moment of inertia considerations, can yield information about dimensions, physical states, and mass distributions of structures within the Earth. But for elemental compositions, one must rely upon implications derived from meteorites.

The constancy in isotopic compositions of most of the elements of the Earth, the Moon, and the meteorites indicates formation from primordial matter of common origin. Primordial elemental composition is yet manifest and determinable to a great extent in the photosphere of the Sun. The less volatile rock-forming elements, present in the outer regions of the Sun, occur in nearly the same relative proportions as in chondritic meteorites. But chondrites differ from one another in their respective proportions of major elements[11,12], in their states of oxidation[13,14], mineral assemblages[15], and oxygen isotopic compositions[16] and, accordingly, are grouped into three distinct classes: *enstatite*, *carbonaceous* and *ordinary*. Virtually all approaches to whole-Earth composition are based upon the idea that the Earth is similar in composition to a chondrite meteorite. A major problem within the Earth sciences began more than six decades ago with the wrong choice of chondrite type as being representative of the Earth.

When earthquake-waves enter and leave the core, they change speed and direction. Consequently, there is a region, called the shadow zone, where earthquake-waves should not be detectable. But in the early 1930s, earthquake-waves were in fact detected in the shadow zone. Inge Lehmann[17] discovered the inner core by showing that a small solid object, within the fluid core, could cause earthquake-waves to be reflected into the shadow zone.

## The Contradiction

Four years after Inge Lehmann discovered the inner core, Francis Birch[18] pronounced its composition to be partially crystallized nickel-iron metal. It is important to understand Birch's logic, which seemed reasonable within the framework of knowledge at the time, but that ultimately set the Earth science community along an incorrect progression of development.

Birch envisioned the Earth to be like an *ordinary* chondrite, the most common type of meteorite observed to fall to Earth. He ignored the rare, oxygen-rich *carbonaceous* chondrites, which contain little or no iron metal, and he ignored the rare oxygen-poor *enstatite* chondrites, which contain some minerals, such as oldhamite (CaS), that are not found in the surface regions of the Earth.

Birch thought that nickel and iron were always alloyed in meteorites and he knew that the total mass of all elements heavier than nickel was too little to comprise a mass as large as the inner core. Birch therefore assumed that the inner core was nickel-iron metal that had begun to crystallize from the melt. That assumption, which underlies much geophysical and geochemical development over the past six decades, is unfounded.

From discoveries made in the 1960s, I realized a different possibility for the composition of the Earth's inner core, which I published in the *Proceedings of the Royal Society of London* in 1979[19]. The abstract in its entirety states: "From observations of nature the suggestion is made that the inner core of the Earth consists not of nickel-iron metal but of nickel silicide". Whereas Birch had thought that nickel and iron were always alloyed in chondrites, I realized that elemental silicon, found in the metal of *enstatite* chondrites[20], under appropriate conditions could cause nickel to precipitate as nickel silicide, an intermetallic compound of nickel and silicon, like the mineral *perryite*, which had been discovered in *enstatite* chondrites[21].

Significantly, a fully crystallized inner core of nickel silicide would constitute a mass virtually identical to the observed mass of the inner core; no such predictability exists for Birch's concept of a partially crystallized nickel-iron metal inner core.

In ethical science, when a contradiction arises to an important concept, the matter should be discussed and debated; experiments and/or theoretical considerations should be made. If the contradiction is found to be in error, it should be refuted, ideally in the journal of original publication; otherwise, it should be acknowledged. The response of the geo-science community to my concept of a nickel silicide inner core was simply to ignore the work for more than two decades, as can be readily verified from Science Citation Index Expanded™ searches (Table 1). In stark contrast, Inge Lehmann, wrote to me, "I admire the precision of your reasoning based upon available information, and I congratulate you on the highly important result you have obtained."

## Oxygen Rules

Only five elements (Fe, Mg, Si, S and O) constitute about 95% of the mass of each chondrite meteorite and, by implication, about 95% of the mass of each of the terrestrial planets. Four of those elements (Fe, Mg, Si and S) occur in chondrites in about the same relative proportion as they occur in the outer regions of the Sun, to within a factor of two. Oxygen is the exception, being about 8 times more abundant in the photosphere of the Sun than the sum of the other four elements of either chondrite shown in Table 2. The high relative abundance of oxygen in solar matter poses a serious limitation on the nature of primordial condensates from that medium. The oxidation states of chondritic matter and, by implication, the oxidation states of the terrestrial planets are set by the nature of those condensates and the circumstances of the separation of those condensates from their primordial gases.

The ordinary chondrites comprise about 80% of the meteorites that are observed falling to Earth and, in terms of the 5 major elements, consist principally of the following minerals: olivine [(Mg, Fe)$_2$SiO$_4$], pyroxene [(Mg, Fe)SiO$_3$], troilite [FeS], and metal [Fe]. Considerable confusion has arisen within the Earth sciences by the promulgation of models during the 1970s that incorrectly assumed that this mineral assemblage formed as condensate from a gas of solar composition[22].

Hans Suess and I[23] demonstrated from thermodynamic considerations that the oxidized iron content of the silicates of ordinary chondrites was consistent, not with formation from solar matter, but instead with their formation from a gas phase depleted in hydrogen by a factor of about 1000 relative to solar composition. Subsequently, I[24] showed that oxygen depletion, relative to solar matter, was also required, otherwise essentially all of the elements would be observed combined with oxygen as they are in the C1 or CI carbonaceous chondrites. Moreover, I showed that if the mineral assemblage characteristic of ordinary chondrites could exist in equilibrium with a gas of solar composition, it is at most only at a single low temperature, if at all. Such a mineral assemblage, therefore, cannot legitimately be assumed to be a primary building component of the Earth. Instead, the ordinary chondrite meteorites appear to have formed from a mixture of two components, re-evaporated after separation from solar gases, one component being an oxidized primitive matter like C1 chondrites, the other being a partially differentiated planetary component from enstatite-chondrite-like matter[25].

Even though the scientific underpinnings of the so-called Equilibrium Condensation Model were shown untrue[24], the model is still being used as the underlying assumption for other Earth composition models, such as the so-called Bulk Silicate Earth Model. Models such as that are typically constructed by assuming elemental abundances, usually like those of C1 or CI chondrites, for elements having zero or positive valances, distributing those elements in different regions of the Earth on the basis of various geochemical assumptions, and assigning oxygen on the basis of stoichiometry. But there is no evidence that nature ever obliged those assumptions and it is not scientifically legitimate to use the Equilibrium Condensation Model as justification.

## Earth-Chondrite Relationships

After an inspiring conversation with Inge Lehmann in 1979, I progressed through the following logical exercise: If the inner core is in fact nickel silicide, as I had suggested[19], then the Earth's core must be like the alloy portion of an enstatite chondrite. If the Earth's core is in fact like the alloy portion of an enstatite chondrite, then the Earth's core should be surrounded by a silicate shell like the silicate portion of an enstatite chondrite. This silicate shell, if it exists, should be bounded by a seismic discontinuity, because the silicates of enstatite chondrites have a different and more highly reduced composition than rocks that appear to come from within the Earth's upper mantle. Using the alloy to silicate ratio of the Abee enstatite chondrite and the mass of the Earth's core, by simple ratio proportion I calculated the mass of that silicate shell. From tabulated mass distributions[26], I then found that the radius of that predicted seismic boundary lies within about 1.2% of the radius at the seismic discontinuity that separates the lower mantle from the upper mantle. This logical exercise led me to discover the fundamental quantitative mass ratio relationships connecting the interior parts of the Earth with parts of the Abee enstatite chondrite that are shown in Table 3. Although I published these fundamental mass ratio relationships more than 20 years ago, they have been systematically ignored by those who make models which are based upon assumptions.

Consider the Earth's core as a percentage by mass of the Earth as a whole, about 32.5%. Similarly, consider the percentage by mass of the alloy portion of each chondrite for which data are available[11,31,48], as shown in Figure 1. Note that, if the Earth has a chondritic composition, as widely believed for good reason, then the Earth is in the main like an enstatite chondrite and *not* like an ordinary chondrite. The reason is readily apparent from inspection of the x-axis; the relative oxygen content in chondritic matter primarily determines the relative amount of alloy. The Earth as a whole, and particularly the endo-Earth, the inner 82%, has a state of oxidation like an enstatite chondrite and *not* like an ordinary chondrite.

The identity of the components of the Abee enstatite chondrite with corresponding components of the Earth means that with reasonable confidence one can understand the composition of the Earth's core by understanding the Abee meteorite or one like it. High pressures, such as prevail within the Earth's core, cannot change the state of oxidation of the core. Not surprisingly, $MgSiO_3$, the major silicate of the Abee enstatite chondrite, has been shown to be a stable phase in a perovskite structure at lower mantle pressures[27,28].

The oxidation state determines, not only the relative mass of the core, but the elements the core contains. Highly reduced matter, like that of the Abee enstatite chondrite and the endo-Earth (*i.e.*, the core and lower mantle), was separated from solar gases under conditions that severely limited the oxygen content[29]. As a consequence certain elements, including Si, Mg, Ca, Ti, and U, which would occur entirely as silicate-oxides in ordinary chondrites, occur in part in the alloy portion of the Abee enstatite chondrite and in the Earth's core. Being unable to form oxides, those core-elements compete on the basis of chemical activity and may be accommodated otherwise, for example, as sulfides.

Commercially, to desulfurate steel, calcium (Ca) or magnesium (Mg) is intentionally introduced into the molten alloy to form CaS and MgS, which precipitates at a high temperature and floats to the surface. The expectation within the Earth's core is for CaS and MgS to precipitate at a high temperature and to float to the top of the core; I have suggested that the "islands" of matter at the core mantle boundary[30] consist of such CaS and MgS[4,14].

Whereas the gross features of the endo-Earth appear relatively simple, consistent with the identification of that part being like an enstatite chondrite, the upper mantle displays several seismic discontinuities suggestive of different layers. The challenge is not to make models assuming their compositions but, rather, to identify with certainty the compositions of those layers by discovering fundamental quantitative relationships. If the Earth is chondritic in composition, the upper mantle may be expected to consist mainly of mixtures of the components from the two "primitive" chondritic matter formation reservoirs that yielded the highly oxidized matter like the C1 or CI chondrites and the highly reduced matter like the enstatite chondrites.

## The Tasks Ahead

Because of the fundamental mass ratio relationships connecting the Earth's core and lower mantle to corresponding parts of the Abee enstatite chondrite (Table 2), one may use analytical data for that meteorite, or one like it, to understand the nature of deep-Earth chemistry, particularly the partitioning of elements between the core and lower mantle, which is governed by oxygen fugacity and sulfur fugacity. Regrettably, the Earth science community has systematically failed to exploit a real opportunity to advance scientific understanding. The best and most comprehensive work on the mineralogy and chemical relationships among enstatite chondrites[31] was published in 1968, at a time when the electron microprobe was not yet fully perfected. The best data on the distribution of actinide elements among components of the Abee meteorite[32] was published in 1982 with thorium data so incomplete as to be virtually useless. Imagine the benefit to be realized from precise and thorough investigations of enstatite chondrites using current, state-of-the-art technology.

Vast resources have been expended for high-pressure diamond-anvil laboratory experiments. Regrettably the results to date are of little relevance for the Earth's interior, being focused mainly on the phase relationships of iron metal. What is needed is to elucidate the phase relations in the system Fe-Ni-S-Si-Mg-Ca at core pressures and temperatures, focusing on compositions near those shown in Table 4. In particular, it is important to reveal the $p$-$T$ conditions under which nickel silicide will precipitate, as these data may place constraints on the temperature at the outer boundary of the inner core. It is important to determine the range of possible compositions of nickel silicide and the physical properties of each[33].

From the standpoint of geo-antineutrino investigations, it is vital to determine the partitioning of naturally occurring radioactive elements between the Earth's core and mantle in the system Fe-Si-Mg-S-O for compositions near that of the Abee enstatite chondrite. Such investigations should go hand-in-hand with state-of-the-art mineralogical and chemical investigations of the enstatite chondrites. Simply showing, for example, that an element like potassium reacts with iron at core *p-T* is without meaning, as the oxygen fugacity and sulfur fugacity are dominant considerations. At present the best quantitative data indicates that about 80% of the uranium of the Abee enstatite chondrite occurs in the alloy portion, implying that about 64% of the Earth's uranium resides within the core. Although there are some observations that suggest thorium and potassium may also occur to some extent within the alloy portion[31,32], one cannot say with any certainty how much. These relative quantities can be determined, but should not be assumed.

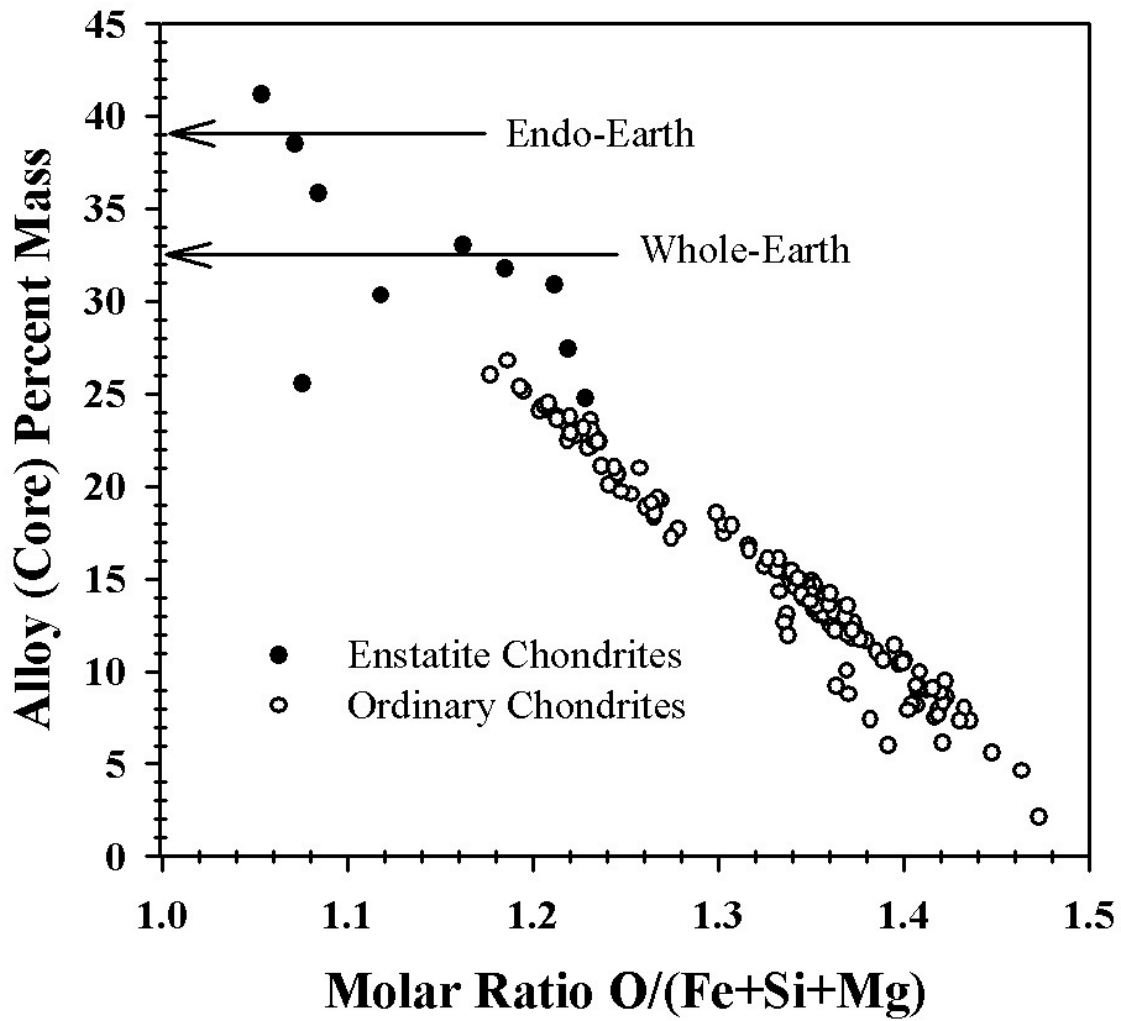

**Figure 1**. The percent mass of the alloy component of each of 9 enstatite chondrites and 157 ordinary chondrites. This figure clearly shows that, if the Earth is chondritic in composition, the Earth as a whole, and especially the endo-Earth, is like an enstatite chondrite and *not* like an ordinary chondrite. The reason is clear from the abscissa which shows the molar ratio of oxygen to the three major elements with which it combines in enstatite chondrites and in ordinary chondrites. This figure also clearly shows that, if the Earth is chondritic in composition, the Earth as a whole, and especially the endo-Earth, has a state of oxidation like an enstatite chondrite and *not* like an ordinary chondrite.

**Table 1**. Citations (except mine) of my major scientific papers on deep-Earth composition and the nuclear georeactor from Science Citation Index Expanded[TM] searches of about 5,900 journals conducted May 13, 2003.

| Year | Scientific Papers | Citations |
|---|---|---|
| 1979 | The nickel silicide inner core of the Earth[19] | four[34-37] |
| 1980 | The chemical composition of the interior shells of the Earth[38] | three[34,39,40] |
| 1982 | The object at the centre of the Earth[41] | none |
| 1993 | Feasibility of a nuclear fission reactor as the energy source for the geomagnetic field[4] | two[36,42] |
| 1994 | Planetary and protostellar nuclear fission: Implications for planetary change, stellar ignition and dark matter[43] | none |
| 1996 | Sub-structure of the inner core of the Earth[14] | two[36,44] |
| 1998 | Examining the overlooked implications of natural nuclear reactors[45] | none |
| 1998 | Composition of the deep interior of the Earth: divergent geophysical development with fundamentally different geophysical implications[33] | none |
| 2001 | Deep-earth reactor: nuclear fission, helium, and the geomagnetic field[6] | two[36,46] |
| 2003 | Nuclear georeactor origin of oceanic basalt $^3$He/$^4$He, evidence, and implications[5] | too soon |

Notes: Only nine papers cited the above ten papers spanning twenty-five years. Two of the nine were published in *Current Science*. Citations in[37] were at the suggestion of Hans E. Suess and in[34] at the suggestion of Hatten S. Yoder, Jr. Citation in[35] was pejorative and without basis.

Table 2. Comparison of molar (atom) abundance ratios, normalized to Fe, of the five major elements in the highly oxidized Orgueil carbonaceous chondrite with those of the highly reduced Abee enstatite chondrite and with corresponding elements in the photosphere of the Sun[47,48].

| Element Ratio | Orgueil Chondrite | Abee Chondrite | Sun |
| --- | --- | --- | --- |
| Fe/Fe | 1.00 | 1.00 | 1.00 |
| Si/Fe | 1.08 | 1.11 | 1.20 |
| Mg/Fe | 1.19 | 0.82 | 1.12 |
| S/Fe | 0.56 | 0.34 | 0.47 |
| O/Fe | 3.18 | 2.86 | 31.3 |

**Table 3**. Fundamental mass ratio comparison between the endo-Earth (core plus lower mantle) and the Abee enstatite chondrite[41].

| Fundamental Earth Ratio | Earth Ratio Value | Abee Ratio Value |
| --- | --- | --- |
| lower mantle mass to total core mass | 1.49 | 1.43 |
| inner core mass to total core mass | 0.052 | *theoretical* 0.052 if $Ni_3Si$ 0.057 if $Ni_2Si$ |
| inner core mass to (lower mantle+core) mass | 0.021 | 0.021 |
| core mean atomic mass | 48 | 48 |
| core mean atomic number | 23 | 23 |

**Table 4**. Estimated masses (x $10^{24}$ kg) of the six most abundant elements in the (whole) Earth's core by analogy to corresponding elements in the alloy portion of the Abee enstatite chondrite.

| Element | Herndon (1980, 1982)[38,41] | Herndon (1993)[4] |
|---|---|---|
| Mg | 0.0475 | 0.0389 |
| Si | 0.0326 | 0.0376 |
| Ca | 0.0184 | 0.0178 |
| S | 0.284 | 0.285 |
| Fe | 1.45 | 1.46 |
| Ni | 0.0831 | 0.0871 |